\documentclass[preprint,preprintnumbers,amsmath,amssymb]{revtex4}
 \usepackage{dcolumn}
 \usepackage{bm}

\begin{document}

 \preprint{Submited to JMP}

 \title{ Inner topological structure of Hopf invariant }

 \author{Ji-rong REN }
 \author{Ran LI}
 \thanks{Corresponding author.Email: liran05@st.lzu.edu.cn}
 \author{Yi-shi DUAN}
 \affiliation{Institute of Theoretical Physics,  Lanzhou University, Lanzhou, 730000, China}

 \begin{abstract}
 In light of $\phi$-mapping topological current theory, the inner
 topological
 structure of Hopf invariant is investigated. It is revealed that
 Hopf invariant is just the winding number of Gauss
 mapping. According to the inner structure of topological current, a
 precise expression for Hopf invariant is also presented. It is the
 total sum of all the self-linking and all the linking numbers of the knot family.
 \end{abstract}

 \maketitle

 \section{Introduction}

 Results from the pure mathematical literature are of little
 difficulty for physicists and can be made accessible to physicists
 by introducing them in common physical methods. In this paper,
 we discuss an object from algebraic topology, Hopf
 invariant, and reveal the inner topological structure of Hopf
 invariant
 in terms of the so-called $\phi$-mapping topological
 current theory.

 Hopf\cite{hopf} studied the third homotopy group of the 2-sphere and showed
 that this group $\pi_3(S^2)$ is non-trivial; later Hopf invented more
 non-trivial fibration and finally
 obtained a series of invariants $\pi_{2n-1}(S^n)$ (where $n$ is positive integer) that now bear his name. In this
 paper, we will stick strictly to the Hopf map from 3-sphere(denoted as $S^3$) to
 2-sphere(denoted as $S^2$) and the invariant $\pi_3(S^2)$ related to this map
 which can be expressed in the integral form\cite{whitehead}
 \begin{equation}
 H=\int_{S^3}\omega\wedge d\omega \;\;,\nonumber
 \end{equation}
 where $\omega$ is a 1-form on $S^3$. Hopf invariant is
 independent of the choice of 1-form $\omega$, which leads to the
 normalization of $\omega$.

 Hopf invariant is
 an important topological invariant in mathematics and have many applications
 not only in condensed matter physics but also in high energy
 physics and field theory. A review article about Hopf fibration is appeared in
 \cite{hopffibration}
 where the author pointed out Hopf fibration occurs in at
 least seven different situation in theoretical physics in various
 guises.
 Hopf invariant $\pi_3(S^2)$ have deep relationships with the Abelian
 Chern-Simons action\cite{witten,chern} in gauge field theory,  self-helicity in
 magnetohydrodynamics(MHD)\cite{helicity} and
 Faddeev-Niemi knot quantum number in Faddeev's model\cite{faddeev}.
 As revealed in this paper,
 Hopf invariant is an important topological invariant to
 describe the topological characteristics of the knot
 family.
 Knotlike configurations as string structures of finite energy(finite action)
 appear in a variety of physical, chemical, and biological scenarios,
 including the structure of elementary particles\cite{faddeev,elementary}, early
 universe cosmology\cite{cosmology}, Bose-Einstein condensation
 \cite{BEcondensation}, polymer folding\cite{polymer}, and DNA replication, transcription, and
 recombination\cite{dna}, and have been taken more and more
 attentions to.

 The $\phi$-mapping topological current theory\cite{duanpfimap}
 proposed by Prof.Duan is a powerful tool not only in
 studying the topological objects in physics, including vortex lines and
 monopoles in BEC\cite{bec}, superfluid\cite{superfulid}
 and superconductivity\cite{superconductor},
 but also in discussing the topological objects in mathematics, for example,
 the inner topological structure of
 Gauss-Bonnet-Chern theorem\cite{gbc}, the second Chern
 characteristic class\cite{chernclass} and homotopy group. In this paper, in light of
 $\phi$-mapping topological current theory, the inner topological structure of Hopf invariant is
 discussed in detail.

 The main research interests of this paper are in the area of
 the inner topological structure of Hopf invariant and the precise expression
 for Hopf invariant.
 This paper is arranged as follows. In Sec.II, some basic
 mathematical ideas, including Hopf map, Hopf fibration and Hopf
 invariant, are briefly presented. By using the
 spinor representation of Hopf map, the inner topological structure of Hopf invariant $\pi_3(S^2)$ is
 discussed in detail
 which is revealed that Hopf invariant is the
 winding number of Gauss mapping $S^3\rightarrow S^3$.
 In Sec.III, in light of $\phi$-mapping topological current theory, an conserved topological current is introduced and its inner structure
 is also presented. In Sec.IV, a precise
 expression for Hopf invariant is obtained which is revealed that it is
 the total sum of all the self-linking and all the linking
 numbers of the knot family. A brief conclusion and
 prospect are appeared in the last section.

 \section{inner topological structure Of Hopf Invariant}

 In this section, we firstly introduce some basic mathematical
 ideas, including Hopf map,
  Hopf fibration and Hopf
 invariant. Then, in light of $\phi$-mapping topological current theory,
 the inner topological structure of Hopf invariant $\pi_3(S^2)$ is
 revealed that Hopf invariant is just the
 winding number of Gauss mapping $S^3\rightarrow S^3$.

 The Hopf map $f: S^3\rightarrow S^2$ arises in many contexts, and can be generalized to
 the map $S^{2n-1}\rightarrow S^n$, where $n$ is an integer. There are several descriptions of the Hopf
 map. Here, the spinor representation is expected to discuss the
 inner topological structure of Hopf invariant. So, firstly, we
 want to construct the spinor representation of Hopf map.
 Consider two complex scalar field $z^1$ and $z^2$ satisfying the
 condition
 \begin{eqnarray}
 |z^1|^2+|z^2|^2=1,\nonumber
 \end{eqnarray}
 form which the two-component normalized spinor $z$ can be introduced like this
 \begin{equation}\label{}
 z=
 \left(
   \begin{array}{c}
     z^1 \\
     z^2 \\
   \end{array}
 \right)\;\;.\nonumber
 \end{equation}
 The normalized condition denotes that $z\in S^3$.
 In fact, it is easy to exhibit the basic Hopf map $S^3\rightarrow
 S^2$.
 Generally, Hopf map can be represented by
 the spinor representation\cite{wilczek,bott}
 \begin{equation}
 m^a(x)=z^{\dagger}(x)\sigma^a z(x), \;\;a=1, 2, 3,\;\;x\in R^3
 \end{equation}
 in which  $\sigma^a$ is Pauli matrix. Now, we will interpret why Eq.(1) is the Hopf map $S^3\rightarrow S^2$.
 Physicists should be familiar with this fact from the elementary
 discussions of the Pauli matrices $\sigma^a$.
 Noticing that Pauli matrix elements satisfy the formula
 \begin{equation}
 \sigma^a_{\alpha\beta}\sigma^a_{\alpha'\beta'}=2\delta_{\alpha\beta'}\delta_{\alpha'\beta}
 -\delta_{\alpha\beta}\delta_{\alpha'\beta'}\;\;,\nonumber
 \end{equation}
 one have $m^am^a=1$
 which denotes that $m^a\in S^2$.
 The definition (1) does really not provide the Hopf map $S^3\rightarrow
 S^2$ because of $x\in R^3$.
 Now, we add the boundary condition
 \begin{equation}
 \vec{m}(x)|_{|x|\rightarrow \infty}\mapsto\vec{m}_0.
 \end{equation}
 where $\vec{m}_0$ is a fixed vector.  It is to say that
 we assume the vector $\vec{m}(x)$ points to the same direction at
 spatial infinity, and therefore the spatial infinity can be
 efficiently contracted to a point, i. e. , $R^3\rightarrow S^3$. Thus, now,  the unit
 vector $\vec{m}$ provides us the Hopf map $S^3\rightarrow S^2$.
 Under the boundary condition (2),  We will not distinguish between $R^3$
 with $S^3$.

 In fact, we can take the Hopf map $\vec{m}$ as a
 projection in the sense that $S^3$ is a principle fibre bundle(Hopf bundle or Hopf fibration)
 over the base space $S^2$ with the structure group $U(1)$. The
 standard fibre of Hopf bundle is $S^1$ which is the inverse image of the
 point of $S^2$ under the Hopf map. In 3-sphere, $S^1$ is
 homeomorphous with knot and the quantity to describe the topology
 of these knots is Hopf invariant which is defined as\cite{bott}
 \begin{equation}
 H=\frac{1}{16\pi^2}\int A\wedge B,
 \end{equation}
 where $A$ is connection 1-form and $B=dA=\frac{1}{2}B_{ij}dx^i\wedge dx^j$ is the Hopf
 curvature 2-form\cite{hopfcuvature}
 and the coefficient $1/16\pi^2$ is the normalized coefficient to ensure that $H$ is an integer. According to Hopf map (1), the
 Hopf curvature can be constructed as
 \begin{equation}
 B_{ij}=\vec{m}\cdot(\partial_i\vec{m}\times\partial_j\vec{m})=\epsilon_{abc}m^a\partial_{i}m^b\partial_{j}m^c.
 \end{equation}
 Noticing that Pauli matrix elements satisfy the formula
 \begin{eqnarray}
 &&\epsilon_{abc}\sigma^{a}_{\alpha\beta}\sigma^{b}_{\alpha'\beta'}\sigma^{c}_{\alpha''\beta''}\nonumber\\
 &&=-2i(\delta_{\alpha\beta'}\delta_{\alpha'\beta''}\delta_{\alpha''\beta}
 -\delta_{\alpha\beta''}\delta_{\alpha''\beta'}\delta_{\alpha'\beta}),
 \nonumber
 \end{eqnarray}
 one can arrive at the following expression for Hopf curvature after
 some algebra
 \begin{equation}
 B_{ij}=-2i(\partial_{i}z^\dagger\partial_{j}z-\partial_{j}z^\dagger\partial_{i}z).\nonumber
 \end{equation}
 Then obviously one can get
 \begin{equation}
 A_i=-2iz^\dagger\partial_{i}z.
 \end{equation}
 which is just the canonical connection\cite{hopffibration} of Hopf bundle.
 In the $\phi-$mapping theory\cite{duanpfimap}, since the spinor
 field $z$ is the fundamental field which is essential to the
 topology properties of manifold itself, the canonical connection
 (5) just reveals the inner structure of Hopf bundle's connection.
 It is easy to see that $A_i$ is in the form of U(1) gauge potential and
 Hopf invariant is unchanged under the gauge transformation
 \begin{equation}
 A'_i=A_i+\partial_i\psi,\nonumber
 \end{equation}
 where $\psi$ is an arbitrary complex scaler function. The invariance presents the $U(1)$ symmetry
 of Hopf fibration and is very important
 to the following discussion.
 Depending on the $U(1)$ invariance of Hopf invariant, we can select
 Columb gauge condition as done in
 classical electrodynamics, i.e. impose the condition
 $\partial_iA_i=0$ in Sec.IV without losing generality.

 In terms of the canonical
 connection form (5),Hopf invariant can be written as
 \begin{equation}
 H=\frac{1}{32\pi^2}\int\epsilon^{ijk}A_iB_{jk}d^3x\\=-\frac{1}{4\pi^2}\int\epsilon^{ijk}z^\dagger\partial_i
 z\partial_j z^\dagger \partial_k z d^3x.\nonumber
 \end{equation}
 Since the spinor field $z$ is the fundamental field on
 manifold and just describes the topological property of the manifold
 itself, the above expression is obviously more direct
 in the study of Hopf invariant.
 The normalized two-component spinor $z$ can be expressed by
  \begin{equation}\label{}
 z=
 \left(
   \begin{array}{c}
     l^0+il^1 \\
     l^2+il^3 \\
   \end{array}
 \right)\;\;,
 \end{equation}
 where $l^a(a=0, 1, 2, 3)$ is a real unit vector.
 In the $\phi$-mapping theory\cite{duanpfimap} the unit vector
 $l^a$ should be further determined by the smooth vectors $\varphi^a$, i.e.
 \begin{equation}
 l^a=\frac{\varphi^a}{\|\varphi\|}\;\;,\;\;\;\;\|\varphi\|=\varphi^a \varphi^a .
 \end{equation}

 Substituting the expression (6) of
 the two-component spinor $z$ into Hopf invariant, one can get
 \begin{equation}
 H=\frac{1}{12\pi^2}\int\epsilon_{abcd} \epsilon^{ijk}l^a\partial_i
 l^b\partial_j l^c \partial_k l^d d^3x.
 \end{equation}
 One can see that the integrated function in the right side is the unit
 surface element of $S^3$ which implies that Hopf invariant is just the
 winding number of Gauss mapping $S^3\rightarrow S^3$. Using
 $\phi$-mapping method the inner topological structure can be
 studied.
 While $S^3$ can be viewed as the infinite
 boundary of $R^4$, using Stokes theorem one can arrive at
 \begin{equation}
 H=\frac{1}{12\pi^2}\int_{R^4}\epsilon_{abcd}
 \epsilon^{\mu\nu\lambda\rho}\partial_{\mu}l^a\partial_{\nu}
 l^b\partial_{\lambda} l^c
 \partial_{\rho} l^d d^4x,\nonumber
 \end{equation}
 where $x^{\mu}$ is the coordinate of $R^4$.  According to
 $\phi$-mapping topological theory\cite{duanpfimap}, one can use Eq.(7)
 and the Green function relation in $\phi$-space
 \begin{equation}
 \frac{\partial^2}{\partial\varphi^a\partial\varphi^a}(\frac{1}{\|\varphi\|})=-4\pi^2\delta^4(\vec{\varphi})\nonumber
 \end{equation}
 to express Hopf invariant in the following form
 \begin{equation}
 H=\int_{R^4}\delta^4(\vec{\varphi})D(\frac{\varphi}{x})d^4x,
 \end{equation}
 where $D(\frac{\varphi}{x})=(1/4!)\epsilon_{abcd}\epsilon^{\mu\nu\lambda\rho}
 \partial_{\mu}\varphi^a\partial_{\nu}
 \varphi^b\partial_{\lambda}
 \varphi^c\partial_{\rho}\varphi^d\nonumber$
 is the Jacobian determinant.  Suppose $\varphi^a(x)(a=0, 1, 2, 3)$
 have $m$ isolated zeros at
 $x^\mu=x^\mu_i(i=1, 2, \ldots, m)$.  According to $\delta$-function
 theory\cite{deltafunction}, $\delta^4(\vec{\varphi})$ can be expressed by
 \begin{equation}
 \delta^4(\vec{\varphi})=\sum_{i=1}^{m}\frac{\beta_i\delta^4(x-x_i)}{|D(\frac{\varphi}{x})|_{x=x_i}},\nonumber
 \end{equation}
 where $\beta_i$ is the Hopf index of the $i$th zero. In topology it means that when the point
 $x^\mu$ covers the neighborhood of the zero point $x^\mu_i$ once, the vector
 field $\varphi^a$ covers the corresponding region in $\varphi$-space
 $\beta_i$ times. Substituting the expanding of $\delta^4(\vec{\varphi})$
 into Eq.(9),  one can get
 \begin{equation}
 H=\sum_{i=1}^{m}W_i,
 \end{equation}
 where $W_i=\beta_i\eta_i$ the winding number of Gauss mapping
 $z:S^3\rightarrow S^3$ with $\eta_i=sign[D(\phi/x)]_{x_i^\mu}=\pm 1$ is the Brouwer degree.

 In this section, we have applied the spinor representation of Hopf map to
 calculate the Hopf invariant, and reveal its inner topological
 structure. Although Hopf map $S^3\rightarrow
 S^2$ arises in many contexts, Hopf invariant do not
 depend on the special representation of Hopf map because of the character of topological invariance .
 As discussed in this section, Hopf invariant is just the winding
 number of Gauss map $S^3\rightarrow S^3$ which means that
 Hopf invariant is equal to the homotopy group
 $\pi_3(S^3)$. In another aspect, Hopf invariant is
 characterized by the homotopy group $\pi_3(S^2)$.
 The above considerations lead us to the following crucial relation $\pi_3(S^3)=\pi_3(S^2)=Z$.
 This can be interpreted as following. In fact, the definition
 of $\vec{m}$ involves a two-step map $S^3\rightarrow S^3\rightarrow S^2$.
 The first step is given by the spinor $z:S^3\rightarrow S^3$, and
 the second step is given by the unit vector $\vec{m}$. The first
 step of the map is an Gauss map. We have discussed that
 Hopf invariant is defined as the winding number of the
 first step map $z:S^3\rightarrow S^3$. In algebraic topology, this result
 is obtained using the machinery of the exact homotopy sequences\cite{bott} of
 fibre bundles.

 \section{The Inner Structure Of Topological Current}

 We find that, from the Hopf curvature (4), a conserved
 topological current can be naturally introduced. In this section,
 we mainly study the inner structure of this conserved topological
 current. This section plays a crucial role in establishing the relationship
 between Hopf invariant and linking number of the knot family.

 According to $\phi$-mapping topological current theory,  we can
 define a topological current
 \begin{equation}
 j^i=\frac{1}{8\pi}\epsilon^{ijk}B_{jk}=\frac{1}{8\pi}\epsilon^{ijk}\epsilon_{abc}m^a\partial_{j}m^b\partial_{k}m^c.
 \end{equation}
 It can be proved that
 \begin{equation}
 B_{jk}=\epsilon_{abc}m^a\partial_{j}m^b\partial_{k}m^c=2\epsilon_{ab}\partial_j
 n^a\partial_k n^b,\nonumber
 \end{equation}
 where $n^a(a=1, 2)$ is a two-dimensional vector in the tangent space
 of sphere $S^2$.The above relation is known as Mermin-Ho relation\cite{merminho}.
 The two-dimensional vector $n^a$ is defined as
 \begin{equation}
 n^a=\frac{\phi^a}{\|\phi\|}, \|\phi\|=\phi^a\phi^a, a=1, 2.
 \end{equation}
 where $\vec{\phi}$ is a two-dimensional vector function on $R^3$.
 Then the topological current can be expressed as
 \begin{equation}
 j^i=\frac{1}{4\pi}\epsilon^{ijk}\epsilon_{ab}\partial_j
 n^a\partial_k n^b\\=\delta^2(\vec{\phi}(x))D^i(\frac{\phi}{x}),
 \end{equation}
 where
 $D^i(\frac{\phi}{x})=\frac{1}{2}\epsilon^{ijk}\epsilon_{ab}\partial_j\phi^a\partial_k\phi^b$
 is the Jacobian vector.  This expression of $j^i$ provides an
 important conclusion
 \begin{equation}
 j^i
 \begin{cases}
 =0& \text{if and only if $\vec{\phi}\neq 0$}, \\
 \neq 0& \text{if and only if $\vec{\phi}=0$}.
 \end{cases}
 \nonumber
 \end{equation}
 So it is necessary to study the zero points of $\vec{\phi}$ to
 determine the nonzero solution of $j^i$.  The implicit function
 theory\cite{implicitfunction} show that, under the regular condition
 $D^i(\frac{\phi}{x})\neq 0$, the general solutions of
 \begin{equation}
 \phi^1(x^1, x^2, x^3)=0, \\
 \phi^2(x^1, x^2, x^3)=0,\nonumber
 \end{equation}
 can be expressed as
 \begin{equation}
 x^1=x^1_k(s), x^2=x^2_k(s), x^3=x^3_k(s).\nonumber
 \end{equation}
 which represent $N$ isolated singular strings $L_k(k=1, 2, \ldots, N)$
 with string parameter $s$.

 In $\delta$-function theory\cite{deltafunction},  one can prove that in three
 dimension space
 the expanding of $\delta^2(\vec{\phi})$ is

 \begin{equation}
 \delta^2(\vec{\phi})=\sum_{k=1}^{N}\beta_{k}\int_{L_k}\frac{\delta^3(\vec{x}-\vec{x}_k(s))}{\mid
 D(\frac{\phi}{u})\mid_{\Sigma_k}}ds,
 \end{equation}
 where
 $D(\frac{\phi}{u})=\frac{1}{2}\epsilon^{ij}\epsilon_{ab}\frac{\partial\phi^a}{\partial
 u^i}\frac{\partial\phi^b}{\partial u^j}$ and $\Sigma_k$ is the $k$th
 planar element transverse to $L_k$ with local coordinates
 $(u^1, u^2)$.  The positive integer $\beta_k$ is the Hopf index of
 $\phi$ mapping.  Meanwhile the tangent vector of $L_k$ is given by
 \begin{equation}
 \frac{dx^i}{ds}\mid_{L_k}=\frac{D^i(\frac{\phi}{x})}{D(\frac{\phi}{u})}\mid_{L_k}.\nonumber
 \end{equation}
 Then the inner topological structure of $j^i$ is
 \begin{equation}
 j^i=\sum_{k=1}^N
 W_k\int_{L_k}\frac{dx^i}{ds}\delta^3(\vec{x}-\vec{z_i}(s))ds,
 \end{equation}
 where $W_k=\beta_k\eta_k$ is the winding number of $\vec{\phi}$
 around $L_k$,  with $\eta_k=sgnD(\frac{\phi}{u})\mid_{\Sigma_k}=\pm
 1$ being the Brouwer degree of $\phi$ mapping.

 It can be seen that when these singular strings are closed curves or
 more generally are a family of $N$ knots $\gamma_k(k=1, 2, \ldots, N)$,
 the inner structure of topological current is
 \begin{equation}
 j^i=\sum_{k=1}^N
 W_k\oint_{\gamma_k}\frac{dx^i}{ds}\delta^3(\vec{x}-\vec{z_i}(s))ds,
 \end{equation}
 and Hopf invariant can be written as
 \begin{equation}
 H=\frac{1}{4\pi}\sum_{k=1}^{N}W_k\oint_{\gamma_k}A_idx^i.
 \end{equation}
 From the discussion of Sec.II,  we see that $A_i$ is similar with the vector
 potential in gauge theory.   But the expression (17) indicates that it is
 only for closed strings that Hopf invariant is unchanged under
 the $U(1)$ gauge transformation of $A_i$.  This is the reason why we only consider the
 closed configuration of these singular strings.

 \section{the precise expression for Hopf Invariant}

 One can find that it is useful to
 express $A_i$ in terms of the topological current $j^i$.
 We know that $A_i$ is in the form of $U(1)$ gauge potential in $U(1)$ electromagnetic gauge
 theory. So the topological current $j^i$ is similar with magnetic field
 and the method in classical electrodynamics should be useful.
 Now we can
 introduce a vector as
 \begin{equation}
 C_i=\epsilon_{ijk}\partial_j j^k.
 \end{equation}
 As pointed out in Sec.II, one can impose the condition $\partial_iA_i=0$ which is to
 say that we select
 the Columb gauge in the discussion. It is easy to get
 \begin{equation}
 C_i=-\frac{1}{4\pi}\partial_j\partial_j A_i.\nonumber
 \end{equation}
 This is just the Possion equation. The general solution of the above equation is
 \begin{equation}
 A_i=-\int d^3y\frac{C_i(y)}{|\vec{x}-\vec{y}|}.\nonumber
 \end{equation}
 Then we obtain the crucial
 relation between $A_i$ and the topological current $j^i$
 \begin{equation}
 A_i=-\epsilon_{ijk}\int
 d^3y\frac{\partial_jj^k(y)}{|\vec{x}-\vec{y}|}.
 \end{equation}
 Then the Hopf invariant is
 \begin{eqnarray}
 H&=&\frac{1}{4\pi}\int A_ij^id^3x\nonumber \\
 &=&-\frac{1}{4\pi}\epsilon_{ijk}\int
 d^3x\int d^3y
 j^i(x)\partial_jj^k(y)\frac{1}{|\vec{x}-\vec{y}|}.\nonumber
 \end{eqnarray}
 Because the integral is on the total space, under the boundary condition (2)
 the topological current vanishes naturally on the boundary.
 Then one can get
 \begin{eqnarray}
 H=\frac{1}{4\pi}\epsilon_{ijk}\int d^3x\int d^3y
 j^i(x)j^j(y)\partial_k\frac{1}{|\vec{x}-\vec{y}|}.\nonumber
 \end{eqnarray}
 Substituting the inner structure of topological current (16)
 into the above equation, one can get
 \begin{equation}
 H=\frac{1}{4\pi}\sum_{m,
 n=1}^{N}W_mW_n\epsilon_{ijk}\oint_{\gamma_m}dx^i\oint_{\gamma_n}dy^j\partial_k\frac{1}{|\vec{x}-\vec{y}|}.
 \end{equation}
 One should notice that the above equation includes two
 cases:(1)$\gamma_m$ and $\gamma_n$ are two different knots($m\neq
 n$).  Noticing that the Gauss linking number is defined as
 \begin{equation}
 {\cal{L}}k(\gamma_m,
 \gamma_n)=\frac{1}{4\pi}\epsilon_{ijk}\oint_{\gamma_m}dx^i\oint_{\gamma_n}dy^j\partial_k\frac{1}{|\vec{x}-\vec{y}|},
 \end{equation}
 we can get a portion of Hopf invariant
 \begin{equation}
 H^{(1)}=\sum_{m, n=1, m\neq n}^{N}W_mW_n{\cal{L}}k(\gamma_m,
 \gamma_n).\nonumber
 \end{equation}
 (2)$\gamma_m$ and $\gamma_n$ are the same knots. It can be proved
 that
 \begin{equation}
 {\cal{L}}k(\gamma_m,
 \gamma_n)={\cal{SL}}(\gamma_m)={\cal{T}}w(\gamma_m)+{\cal{W}}r(\gamma_m),\nonumber
 \end{equation}
 where ${\cal{SL}}(\gamma_m)$ is the self-linking number of
 $\gamma_m$. This formula is well known as White
 formula\cite{whiteformula}
 with ${\cal{T}}w(\gamma_m)$ and ${\cal{W}}r(\gamma_m)$
 being the twisting number and writhing number of $\gamma_m$
 respectively.  So we get another portion of Hopf invariant
 \begin{equation}
 H^{(2)}=\sum_{m=1}^{N}W_m^2{\cal{SL}}(\gamma_m).\nonumber
 \end{equation}
 Finally,  we add the two parts up and get an important expression for
 Hopf invariant
 \begin{equation}
 H=\sum_{m=1}^{N}W_k^2{\cal{SL}}(\gamma_m)+\sum_{m, n=1, m\neq
 n}^{N}W_mW_n{\cal{L}}k(\gamma_m, \gamma_n).
 \end{equation}
 This precise expression reveals the relationship between the Hopf
 invariant and the self-linking and linking numbers of
 $N$ knots family. Since the self-linking
 and linking numbers are both invariant characteristic numbers of the
 knotlike closed curves in topology, the Hopf invariant is an important invariant
 required to describe the knotlike configurations in physics.

 \section{Conclusion}

 In this paper, in light of $\phi$-mapping topological current theory, the inner
 structure of Hopf invariant is studied in detail. It is revealed that
 Hopf invariant is just the winding number of Gauss
 mapping of 3-sphere.
 we also introduce a conserved topological current from which
 a family of knots can be deduced naturally.
 According to the inner structure of topological current, a
 precise expression for Hopf invariant is presented. It is the
 total sum of all the self-linking and all the linking numbers of the
 knot family. A trivial generation of Hopf invariant from $\pi_3(S^2)$ to
 $\pi_{2n-1}(S^n)$(where $n$ is positive integer) is valuable for further study.

 At last, to complete this paper, a final remark is necessary.
 An interesting question arise that the magnetic field
 may be described by the topological current. Because the magnetic
 force lines are closed in the case that there is no magnetic
 monopole existed in the spacetime, the topology and geometry of magnetic field
 can be studied by the $\phi$-mapping topological current theory.
 Recently, Faddeev and Niemi have constructed a model\cite{magneticgeometry} to describe
 the knotted and
 linked configuration of electrical neutral plasmas, which may be
 provide us a theoretical framework to study this question.

 \section*{ACKNOWLEDGEMENT}

 This work was supported by the National Natural Science Foundation
 of China.

 \end{document}